\documentclass[superscriptaddress,preprint,layout=onecolumn]{revtex4-2}
\usepackage{graphicx}
\usepackage{amssymb}
\usepackage{epstopdf}
\usepackage{amsmath,dsfont, commath}
\usepackage{bm}
\usepackage{braket}
\usepackage{changes}
\usepackage{siunitx}

\newcommand{\be}{\begin{equation}}
	\newcommand{\ee}{\end{equation}}
\newcommand{\bea}{\begin{eqnarray}}
	\newcommand{\eea}{\end{eqnarray}}

\newcommand{\ba}{\begin{array}}
	\newcommand{\ea}{\end{array}}

\newcommand{\bl}{\begin{flalign}}
	\newcommand{\enl}{\end{flalign}}

\newcommand{\mc}[1]{\mathcal{#1}}

\newcommand{\eq}[1]{Eq. \eqref{#1}}

\renewcommand{\sec}[1]{Sec. \ref{#1}}

\newcommand{\half}{\frac{1}{2}}

\newcommand{\proj}[1]{\ket{#1}\bra{#1}}

\renewcommand{\bf}[1]{\mathbf{#1}}
\newcommand{\grad}{\nabla}

\usepackage[capitalize]{cleveref}

\begin{document}
	
	\title{Density matrix renormalization group in the discrete variable representation basis}
\author{Bing Gu}

\email{gubing@westlake.edu.cn}
\affiliation{Department of Chemistry and Department of Physics, Westlake University, Hangzhou, Zhejiang 310030, China}
\affiliation{Institute of Natural Sciences, Westlake Institute for Advanced Study, Hangzhou, Zhejiang 310024, China}

\begin{abstract}

We present a numerical implementation of the density matrix renormalization group (DMRG)
using the discrete variable representation (DVR) basis set.
One main advantage of using the local DVR basis sets is that the computations of one-electron integral and two-electron repulsion integrals are drastically simplified.
For comparison, we further implemented DVR complete active space configuration interaction (CASCI) using canonical molecular orbitals.  
These methods are applied to a one-dimensional pseudo-hydrogen chain under screened Coulomb potential. The DMRG ground state energy agrees with CASCI up to 0.1 \si{\milli\hartree} with a very small number of bond dimensions. 
\end{abstract}
\maketitle

\section{Introduction}
Electronic structure is the cornerstone of modern chemistry and is now routinely performed for molecules and materials to understand their physicochemical properties and spectroscopy \cite{lischka2018}. 
Most of the electronic structure codes employ Gaussian-type orbitals as the atomic integrals can be analytically computed. 
The two-electron integrals are fundamental components required for any electronic structure computations \cite{helgaker2000}, e.g., to construct the Fock matrix in the Hartree-Fock (HF) method. 
The evaluation of electron-repulsion integrals is computationally demanding as it scales $N^4$ with the size of the basis set $N$. 
Real-space grids provide an alternative universal basis set for electronic structure computations \cite{white1989, chelikowsky1994, lippert1997, liu2003, rakhuba2016, mortensen2024, kim2015, enkovaara2010}.

The discrete variable representation (DVR) basis set using both a finite number of basis functions and a set of grid points, has been widely used in solving the molecular vibrational eigenstates problems \cite{light2000, gu2023b, gu2024a, zhu2024, littlejohn2002, colbert1992} as it simplifies the computations of both kinetic energy and potential energy operator matrix elements.   
However, much less is explored to use DVR basis sets for electronic structure computations, especially in advanced multi-configurational  methods \cite{jones2016, liu2003}. 

Here we present an implementation of the density matrix renormalization group (DMRG) using DVR basis sets. 
DMRG is a numerical method developed to study quantum many-body systems, particularly in one-dimensional systems. Originally introduced by White in 1992 \cite{white1992}, DMRG has become one of the most powerful techniques in condensed matter physics and, more recently, in quantum chemistry for calculating ground states and low-energy excited states of quantum systems \cite{ma2022, chan2011, olivares-amaya2015}. DMRG represents many-body states using matrix product states or tensor networks, composed of interconnected tensors with a restricted entanglement. 
While conventionally the quantum chemistry DMRG uses the canonical molecular orbitals as sites, using the localized basis can enhance its performance \cite{stoudenmire2017}.   Besides locality, 
one main advantage of using a DVR basis set is the electron-repulsion integral scales, instead of $\mc{O}(N^4)$ using Gaussian-type orbitals, linearly with the number of basis functions $\mc{O}(N)$.  

For comparison, we also implement the  complete active space configuration interaction (CASCI)  method employing DVR basis sets, with the complete active space defined by the HF canonical molecular orbitals. 
CASCI includes all possible electronic configurations (Slater determinants) by distributing electrons among the active orbitals\cite{levine2021}. Unlike the complete active space self-consistent field, the active orbitals are predetermined, e.g. canonical HF molecular orbitals. 
 All codes are implemented in our in-house {Python}-based package \textsc{PyQED}.
An illustration of the DMRG/DVR and  CASCI/DVR method is shown for a one-dimensional pseudo-hydrogen chain with screened Coulomb potential. 


This paper is organized as follows. In \sec{sec:method}, we present the method of DMRG in a DVR basis set and then describe the implementation details. The application to a one-dimensional pseudo-hydrogen chain model is shown in \sec{sec:model}. We discuss the challenges and future perspectives of DVR basis sets in \sec{sec:summary}.

Atomic units $\hbar = e = m_\text{e} = 1$ are used throughout. 

\section{Method}\label{sec:method}

\subsection{DVR basis set}
The DVR basis sets consist of a projection operator $\hat{P}$, usually defined by a finite number of basis functions $P = \sum_{n=1}^N \proj{u_n}$, and a set of grid points $\set{x_i}$ \cite{littlejohn2002, light2000}.   The DVR basis functions are formerly defined as a basis set satisfying simultaneously orthonormality 
\be \braket{\phi_i|\phi_j} = \delta_{ij} \ee 
 and interpolation properties 
\be \braket{x_j| \phi_k} = w_j^{-1/2} \delta_{jk}.  
\ee 
where $\set{w_j, j =1, \dots, N}$ are the weights of the grid points. 
One example of DVR sets satisfying exactly both properties are 
\be \phi_i(x) = {\frac{1}{\sqrt{\Delta x}}\text{sinc}\del{\pi(x - x_i)/\Delta x}}
\ee  
where $x_i = x_0 +				 i \Delta x, i = 0, \pm 1, \pm 2, \cdots$ and $\Delta x$ is the grid spacing, corresponding to a projection operator of band-limited functions \cite{littlejohn2002}. 
The advantage of a DVR is the highly localized character of the basis functions about the grid points. 
Thus,
only a small number of functions is required to represent a spatially localized electronic orbital.

The DVR basis functions are approximately the eigenstates of the projected position operator $X = \hat{P} \hat{x} \hat{P}$ as 
\be
{X  \ket{\Delta_i} + P\hat{x} Q} \ket{x_i} = x_i \ket{\Delta_i}  
\ee 
where 
$
\ket{\Delta_i} = \hat{P} \ket{x_i} 
$ is the unnormalized DVR basis, $\ket{\phi_i} = w_i^{-1/2} \ket{\Delta_i}$, $w_i = {\braket{\Delta_i|\Delta_i}}$ is the weight of $i$th grid point, and $Q = 1 - P$. In the complete basis set limit, $Q = 0$, thus $\ket{\Delta_i}$ ($\ket{\phi_i}$) becomes exact eigenstates of the position operator. In practice, 
\be
 X  \ket{\Delta_i} \simeq x_i \ket{\Delta_i}  
\ee 
Therefore, a convenient way to construct a DVR basis set is to diagonalize the $X$ matrix.

An important property of  DVR basis sets is that for coordinate-dependent operators $\hat{O}(x)$ (e.g. potential energy operator) can be simply computed  by a diagonal approximation 
\be
O_{ij} = \braket{\phi_i | {O}(\hat{x}) | \phi_j } \approx  \delta_{ij} O(x_j) 
\ee   
For any function $\ket{\psi} = \sum_{i=1}^N c_i \ket{\phi_i}$, the expansion coefficients can be simply obtained by left multiplying $\bra{x_j}$
\be
c_j =  \sqrt{w_j}\psi(x_j).
\ee  
Moreover, the kinetic energy operator matrix elements can be computed exactly, instead of using a finite difference method which requires a dense grid.

	A straightforward way to construct a multidimensional DVR is by direct product.
	For example, for $d = 2$, \be
	{\chi_{\bf i}}(\bf r) = {\phi_{i_1}}(x) {\phi_{i_2}}(y) 
	\ee 
	where $\bf i = (i_1, i_2)$ and DVR grid $\bf r_{\bf i} = \del{x_{i_1}, y_{i_2}}$. 
	
	\subsection{DMRG/DVR}
	Quantum chemistry computations starts with building the one- and two-electron integrals in a chosen basis set. 
With a DVR basis set, the kinetic energy operator $t_{ij} = \braket{\phi_i | -\half \grad^2|\phi_j}$ can be computed exactly, 
the electron-nuclear interaction is diagonal  	
\be  \braket{\phi_i| v_\text{en}(\bf R) | \phi_j} = v_i(\bf R) \delta_{ij}.
\label{eq:ven}
\ee
where 
\be v_{\bf i} = \sum_I -Z_I v_\text{C}(|\bf r_{\bf i} - \bf R_I|)
	\ee  is the electron-nuclear interaction and $Z_I$ is the charge of the $I$th nucleus. 
		In \eq{eq:ven}, we have employed the diagonal approximation for the matrix elements of the electron-nuclear interaction. 
The electron repulsion integral (using chemists notation)
\be
v_{ijkl} \equiv (ij|kl) = \iint \dif \bf r_1 \dif \bf r_2 \phi_i^*(\bf r_1) \phi_j(\bf r_1) v_\text{C}(\bf r_1 - \bf r_2) \phi_k^*(\bf r_2) \phi_l(\bf r_2)  
\ee  
in the DVR basis sets can be simplified to 
\be
\del{il | jk} \approx \delta_{il} \delta_{jk} g_{ik} 
\label{eq:eri} 
\ee
where $g_{ik} = (ii|kk)$. 
\cref{eq:eri} reduces significantly the computational cost of electron-repulsion integral. 
Thus, the electronic Hamiltonian in the second-quanized form is given by 
	\be 
	H(\bf R) = \sum_{i,j} \sum_\sigma \del{ t_{ i j} + \delta_{ij} v_i(\bf R) } c_{\bf i \sigma}^\dag c_{\bf j \sigma}  +  \half \sum_{\sigma, \tau = \set{\alpha, \beta}}\sum_{i,j=1}^{N} g_{ij}(\bf R) c_{\bf i \sigma}^\dag  c_{\bf j \tau}^\dag c_{\bf j \tau} c_{\bf i\sigma} + V_\text{nn}(\bf R)
	\label[type]{eq:h}
	\ee 
	where $V_\text{nn}$ is the nuclear repulsion energy. 

For the DMRG implementation,	it is convenient to rewrite the Hamiltonian as (suppressing the nuclear coordinates)
 \be 
 H = \sum_i \del{t_{ii} + v_i} n_{i} + g_{ii} \del{ n_{i\uparrow} n_{i\downarrow} + n_{i\downarrow} n_{i\uparrow} } + \sum_{i < j} \del{t_{ij} \hat{E}_{ij} + \text{H.c.}} +  \half \sum_{i < j} \del{ g_{ij} n_i n_j + \text{H.c.} }
 \ee 
 where $n_i = \sum_{\sigma = \set{\uparrow, \downarrow}} c_{i\sigma}^\dag c_{i\sigma}$ is the single-site electron number operator,  
$\hat{E}_{ij} = \sum_\sigma c_{i\sigma}^\dagger c_{j\sigma}$, H.c. stands for Hermitian conjugate. 


The DMRG calculation can be  either carried out in the original sweeping algorithm \cite{white1992} or the matrix-product state framework \cite{ma2022}.
The former is employed in the current implementation. The main steps are described below.

 \begin{enumerate}
 	 
 		\item We first create electronic DVR basis sets $\set{\ket{\phi_{j}}, j = 1, 2, \dots, L}$. There are many kinds of DVR sets such as the sinc functions, Gauss-Hermite DVR, and particle-in-a-box eigenstates (i.e., sine functions) \cite{colbert1992, light2000, littlejohn2002}. The optimal choice depends on the specific problem and the boundary conditions. 
 		
 \item The DMRG calculation is initiated by the infinite DMRG algorithm. 
For each step, 
the system  block is enlarged by one more site, the Hamiltonian of  $l+1$  sites is represented in terms of $\set{ \ket{\psi^{0,\dots, l-1}_n} \otimes \ket{\sigma_{l}}, n = 0, 1, \dots, D-1}$. Similarly for the environment block. Build the entire superblock Hamiltonian of $2(l+1)$ sites in terms of 
$\set{ \ket{\psi^{0,\dots, l-1}_n} \otimes \ket{\sigma_{l}}  \otimes \ket{\sigma_{l+1}} \otimes \ket{\psi_m^{l+1,\dots, 2l+1}}   }  $. 
\item 
 Compute the ground state energy and eigenstate of the entire system by e.g. Lanczos algorithm, compute the reduced density matrix followed by Schmidt decomposition (i.e. singular value decomposition), retain the $D$  eigenvectors with largest Schmidt coefficients. The same is applied for the environment block. Then,  the system-environment interaction operators is updated with the new bases.
\item 
When the entire system reaches the desired length $L$ (i.e., number of DVR set), we start the sweeping algorithm. During the sweep, the system block grows at the expense of the environment block, with the total number of sites (orbitals) conserved. 
 \end{enumerate}





Here  symmetry with the conservation of electron number and electron spin is not exploited to speed up the calculation. We fix the number of electrons by an energy penalty   
\be 
H_N = \mu \del{\hat{N} - N}^2 =  \mu \sum_{i, j} n_i n_j - 2 \mu N \sum_i n_i +  \mu N^2 
\ee 
where $\hat{N} = \sum_{i=1}^L n_i$ is the total number operator. 

	\subsection{ Hartree-Fock/DVR}
	For comparision, we also implemented the CASCI using DVR basis set. The active space is defined by the HF orbitals. 
		In the restricted HF theory, the many-electron wave function is approximated by a single Slater determinant (closed shell)
		\be \ket{\Phi_0} = | \psi_1 \bar{\psi}_1 \cdots \psi_{n_\text{e}/2} \bar{\psi}_{n_\text{e}/2} |
		\ee 
		 where $\psi_p (\bar{\psi}_p)$ denotes, respectively, molecular orbitals with spin up (down), $n_\text{e}$ is the number of electrons. The MOs are expanded using the DVR basis set, 
		\be
		\psi_n(\bf r) = \sum_{\bf i} \chi_{\bf i}(\bf r) C_{\bf i, n}  
		\ee 
	The optimal molecular orbitals minimizing the total energy are obtained by diagonalizing the  Fock matrix
	\be
	\bf F[\bf C] = \bf h + 2\bf J - \bf K 
	\ee 
	where 
	\be h_{\bf i \bf j} = \braket{\chi_{\bf i} | -\half \Delta + v_\text{en}(\bf R) | \chi_{\bf j}} = t_{\bf i \bf j} + \delta_{\bf i \bf j} v_{\bf i}
	\label{eq:core}
	\ee is the DVR core Hamiltonian consisting of the kinetic energy and electron-nuclear interaction,  $t_{\bf i \bf j} = \Braket{\chi_{\bf i} | -\half \Delta | \chi_{\bf j}}$ is the DVR kinetic energy matrix elements,  $\Delta$ is the Laplacian in $d$-dimensions,

	The Hartree potential
	\be J_{\bf i \bf j} = \delta_{\bf  i  \bf j} \sum_k v_{\bf i \bf k} D_{\bf k \bf k} \ee  is diagonal in the DVR set,  and the exchange 
	\be 
	K_{\bf i \bf j} = v_{\bf i \bf j} D_{\bf i \bf j}.
	\ee 
	Here $D_{\bf i \bf j} = \braket{\Psi_0| c_{\bf j}^\dag c_{\bf i}  |\Psi_0}$ is the DVR one-electron reduced density matrix, and 	$ v_{\bf i \bf j} = (\bf i \bf i \mathbf{|jj})$ 
	is the electron repulsion. 
	 The HF/DVR calculation scales linearly with the number of basis functions $N$. 
	
\subsection{CASCI by Jordan-Wigner transformation}

	CASCI is a full configuration-interaction computation in a chosen subset of single-electron orbitals, i.e., the complete active space. The orbitals can be the canonical molecular orbitals defined in the HF theory, or natural orbitals defined as the eigenstates of the one-electron reduced density matrix.

Upon the convergence of the self-consistent field cycle,  the electronic Hamiltonian can be  transformed from DVR set to molecular orbitals, 
\be 
H = \sum_{p, q} \sum_{\sigma = \uparrow, \downarrow} h_{pq}c_{p\sigma}^\dag c_{q\sigma} + \sum_{p, q, r, s} \sum_{\sigma, \tau = \uparrow, \downarrow} \half (pq|rs) c_{p\sigma}^\dag c_{r\tau}^\dag c_{s\tau} c_{q\sigma}
\label{eq:hmo}
\ee 
The MO electron repulsion integrals in \eq{eq:hmo} can be easily obtained by 
\be 
(pr|qs) = 
\sum_{\bf i, \bf j} U_{\bf i p} U_{\bf jq} U_{\bf i r} U_{\bf j s} v_{\bf i \bf j} 
\ee 

The CASCI computation can proceed by building a full configuration interaction Hamiltonian within the Slater determinants or configuration state functions corresponding to all possible excitations in the active space \cite{helgaker2014}.  Slater determinants are chosen here as it simplifies the implementation \cite{candanedo2023}, 
\be
\ket{\Psi_\alpha} = \sum_I c^\alpha_I\ket{\Phi_I} .
\ee 
and the configuration intersection coefficients $\bf c^\alpha$ are obtained by diagonalization,
\be
\bf H \bf c^\alpha = E_\alpha \bf c^\alpha .
\ee 

Each electronic configuration, or Slater determinant, is labeled a binary array $B_{p\sigma}^I$, where $p$ labels the orbital, $\sigma = \uparrow, \downarrow$ is the spin index, and $I$ labels the Slater determinant \cite{davidsherrill1999}. The matrix elements between determinants $\braket{\Phi_J|H|\Phi_I}$ are computed by the Slater-Condon rules \cite{slater1929,condon1930}.

We additionally use an alternative approach mapping an interacting electronic model to an interacting spin model by the Jordan-Wigner transformation \cite{jordan1928}. This is widely used to map an electronic structure problem onto a  quantum computer \cite{mcardle2020}.  
For spinful electrons, the transformation reads 
\be 
\begin{split}
c_{\uparrow, j} &\leftrightarrow (-1)^{\sum_{l < j} n_{\uparrow, l} + n_{\downarrow, l}} \sigma_{\uparrow, j}^-, \\ 
c^\dag_{\uparrow, j} &\leftrightarrow (-1)^{\sum_{l < j} n_{\uparrow, l} + n_{\downarrow, l}} \sigma_{\uparrow, j}^+, \\ 
c_{\downarrow, j} &\leftrightarrow (-1)^{\sum_{l < j} n_{\uparrow, l} + n_{\downarrow, l}} (-1)^{n_{\uparrow, j}}\sigma_{\downarrow, j}^-, \\ 
c^\dag_{\downarrow, j} &\leftrightarrow (-1)^{\sum_{l < j} n_{\uparrow, l} + n_{\downarrow, l}}  (-1)^{n_{\uparrow, j}} \sigma_{\downarrow, j}^+
\end{split}
\ee 
$\sigma_{\uparrow, j}^\pm, \sigma_{\downarrow, j}^\pm$ are  the spin-1/2 Pauli matrices at $j$th site. For a single site in the  $\set{ \ket{0}, \ket{\uparrow}, \ket{\downarrow}, \ket{\uparrow\downarrow} }$ basis, 
\be
a_{\uparrow} =\sbr{ 
\begin{array}{cccc}
	0 & 0 & 1 & 0 \\ 
	 0 & 0& 0 &  1 \\ 
	  	 0 & 0& 0 &  0 \\ 
	  	 	 0 & 0& 0 &  0 \\ 
	\end{array} }, ~~~
a_{\downarrow} =\sbr{ 
	\begin{array}{cccc}
		0 & 1 & 0 & 0 \\ 
		0 & 0& 0 &  0 \\ 
		0 & 0& 0 &  -1 \\ 
		0 & 0& 0 &  0 \\ 
\end{array} }
\ee 
 As the creation and annihilation operators always appear in pairs in the Hamiltonian, it is useful to note that $n_{\uparrow, j} \leftrightarrow \half \del{1 + \sigma^z_{\uparrow, j}},  n_{\downarrow, j} \leftrightarrow \half \del{1 + \sigma^z_{\downarrow, j}}$. 

For the Jordan-Wigner transformed Hamiltonian, the many-spin eigenstates can be expressed as 
\be 
\ket{\Psi_\alpha} = \sum_{\bm \sigma} C^{\alpha}_{\sigma_1 \sigma_2 \cdots \sigma_L} \ket{\sigma_1 \sigma_2 \cdots \sigma_L}
\ee 
where $L$ is the number of orbitals in the active space, $\ket{\sigma_l} = \set{ \ket{0}, \ket{\uparrow}, \ket{\downarrow}, \ket{\uparrow\downarrow} }$. The size of Hamiltonian $4^L$ is bigger than the size of the configuration interaction Hamiltonian because only the Slater determinants with a fixed number of electrons and fixed spin (e.g. $\braket{S_z} = 0$) are selected. The eigenstates thus obtained may not be pure spin states.

	\subsection{Frozen Core approximation}
The full configuration interaction method uses all MOs to build Slater determinants.  This is impractical as the computational cost scales exponentially with the number of orbitals. The active space is usually a subset of the MOs with the doubly occupied core orbitals frozen. 
	The energy of the frozen core electrons is the same as a closed-shell molecule 
	\be 
	E_\text{FC} = 2\sum_{f \in \text{FC}} h_{ff} + \sum_{f,g} 2J_{fg} - K_{fg}
	.\ee  
	The interaction between the active space  and frozen core orbitals reads 
	\be 
	V_\text{AS-FC} =  \sum_\sigma \sum_{i, j \in \text{AS}} \sum_{k \in \text{FC}} \del{ 2\del{ij|kk} - \del{ik|kj}} c_{i\sigma}^\dag c_{j\sigma}
	\ee 
	where $i,j$ refers to orbitals in the active space and $k$ the frozen core orbitals.
	The second-quantized Hamiltonian in the active space reads 
	\be 
	 H_\text{AS} = E_\text{FC} + \sum_{i, j} \tilde{h}_{ij} c^\dag_{i \sigma} c_{j \sigma}  + 
	 \sum_{i,j,k,l} \sum_{\sigma, \tau = \uparrow, \downarrow} \half (ij|kl) c_{i \sigma}^\dag c_{k \tau}^\dag c_{l\tau} c_{j \sigma}
	\ee 
	where $\tilde{h}_{ij} = h_{ij} + \sum_{k \in \text{FC}} { 2\del{ij|kk} - \del{ik|kj}}$.

	\section{Application to 1D  chain model}\label{sec:model}
		We consider a one-dimensional pseudo-hydrogen chain model with screened Coulomb interaction \cite{stoudenmire2012}. The  electronic Hamiltonian reads
	\be
	H(\bf z) = \sum_{i=1}^{N_\text{e}} -\half \grad_i^2 + v_\text{ee} + v_\text{en}(\bf z) + v_{nn}(\bf z)
	\ee
	where $\bf z$ is the proton positions along the chain.
	We use the regularized Coulomb interaction
	\be v_\text{C}(r) = \frac{\text{erf}(r)}{r}. 
	\ee 
	This model mimics a single transversal basis limit of  the sliced  basis DMRG  \cite{stoudenmire2017}. 

We use the sine DVR with grid points uniformly distributed in the range of (-15, 15) \AA.  The boundary points are not included as the open boundary condition is imposed.  For the DMRG, there is no need for the HF calculation since we directly use the DVR basis as the sites. For the CASCI, 6 active canonical molecular orbitals with 4 electrons are used. 
%
		The molecular geometry is  $\bf z_0 = [-L/6,  L/6]$, $L = 10$ \AA\ with the first and last protons   at $-L/2$ and $L/2$, respectively. 
		The results for the ground state energy are shown in \cref{tab:energy}. The DMRG energy with $D = 12$ matches the CASCI energy with 12 active orbitals up to $10^{-4}$ \si{\hartree}.

	\section{Conclusion}\label{sec:summary}
	
%


 We have developed an implementation of the DMRG and CASCI using DVR basis set.  The electron repulsion integrals can be easily computed and scales linearly with the number of basis functions. The DMRG and CASCI with DVR basis sets  are applied to a one-dimensional pseudo-hydrogen chain model with screened Coulomb interaction. The DMRG results with a small bond dimension agrees with the CASCI results up to 0.1 \si{\milli\hartree}. 

Our results demonstrate the utility of the DVR basis sets in DMRG and multiconfigurational electronic structure methods. 
	Further developments involve extension to realistic molecules and exploiting symmetry to reduce the computational cost. 
	
	
	\begin{table}[hbpt]
		\begin{center}
			\begin{tabular}{ c c c }
				\hline 
				$L$ &  Method & Energy (\si{\hartree})\\ \hline 
						32 & HF & -1.412298  \\
								32  & CASCI(6, 4) &	-1.423990 \\ 
				32 & CASCI(12, 4) & -1.425417\\
				32 & DMRG ($D=10$) &  -1.424155 \\ 
				32 & DMRG ($D=12$) & -1.425322 \\
				\hline 
			\end{tabular}
		\end{center}
		\caption{Energies of DMRG/DVR and CASCI($n_\text{o}, n_\text{e}$)/DVR with different sizes of basis sets. $D$ is the maximum number of states (bond dimension in the language of matrix product states) retained in DMRG.}
		\label{tab:energy}
	\end{table}

\begin{acknowledgements}
This work is supported by the National Natural Science Foundation of China (Grant No.92356310).
	\end{acknowledgements}

	\appendix

	\section{Implementation of CASCI/DVR}\label{sec:implementation}

	The main steps for implementing the  CASCI  with DVR basis sets and Jordan-Wigner transformation are as follows 

	\begin{enumerate}
		\item We first create electronic DVR basis sets $\set{\ket{\chi_{\bf j}}}$, labeled by ${\bf j} = (n_1, \cdots,n_d)$ where $d$ is the dimensionality of real-space. 

		\item Perform mean-field HF calculation to yield molecular orbitals (MOs)
	\be 
	\ket{\phi_p} = \sum_{\bf j}  \ket{\chi_{\bf j}} U_{\bf j p}
	\ee 
		 and orbital energies $\varepsilon_p$. 
	The DVR to MO transformation is used to transform the creation and annihilation operators, 
	\be 
	c_p^\dag = \sum_{\bf j} U_{\bf j p} c_{\bf j}^\dag, ~~~  c_{\bf j}^\dag = \sum_p [U^\dag]_{p \bf j} c_p^\dag.  
	\ee 

	The HF self-consistent field computations are similar to the standard Gaussian-type orbital-based computations \cite{szabo1996, thijssen1999}, the only difference is that the one-electron and two-electron integrals are straightforwardly evaluated in the DVR basis set. For spin-unrestricted calculations, 
	\be 
	\phi_{p\sigma}(\bf r) = \sum_{\bf i} C^{\sigma}_{\bf i p} {\chi_{\bf i}}(\bf r)  .
	\ee 

		\item Fill electrons according to the Aufbau principle. 
		\item Choose the active space with $L$ MOs.  
		\item Transform the electronic Hamiltonian to the second quantized form in the active space. 

		\item Map the electronic Hamiltonian to a spin model by Jordan-Wigner transformation.  
	 The transformed	 Hamiltonian is highly sparse with size  
	 $4^L$ for spin-unrestricted calculations as each orbital contains four states $\ket{0}, \ket{\uparrow}, \ket{\downarrow}, \ket{\uparrow\downarrow}$.  
	\item		Use Lanczos or Davidson matrix diagonalization method to obtain low-lying eigenenergies $E_\alpha$  and eigenstates $\ket{\Psi_\alpha}$.  When the active space is large $L > 20$, a direct diagonalization  is impractical and more efficient methods such as the density-matrix renormalization group methods  and the iterative configuration intersection can be used \cite{ma2022, white1992, liu2016b}.  

	\end{enumerate}

		\bibliography{../../../qchem,../../../dynamics}
\end{document}